\documentclass[onecolumn,authoryear]{els-mrw} 

\usepackage{amsmath,amssymb,amsfonts,amsthm,makeidx,graphicx}
\usepackage{txfonts}
\usepackage{helvet}

\newcommand{\nat}{nature}
\newcommand{\aj}{AJ}
\newcommand{\apj}{ApJ}
\newcommand{\apjs}{ApJS}

\newcommand{\mnras}{MNRAS}

\newcommand{\araa}{Annu. Rev. Astron. Astrophys.}

\newcommand{\ssr}{Space Sci. Rev.}
\newcommand{\pasa}{PASA}

\newcommand{\msun}{$M_{\odot}$}

\newcommand{\mh}{$M_{H_2}$}
\newcommand{\mhi}{$M_{HI}$}

\newcommand{\mstar}{$M_{\ast}$}

\newcommand{\hi}{{H{\sc i}}}
\newcommand{\hii}{{H{\sc ii}}}
\newcommand{\hmol}{H$_2$}

\newcommand{\tdep}{$t_{dep}({\rm H_2})$}

\newcommand{\fgas}{$f_{H_2}$}
\newcommand{\fhi}{$f_{HI}$}

\begin{document}

\chapter{The Interstellar Medium}

\author[1,2]{Am\'elie Saintonge}

\address[1]{\orgname{University College London}, \orgdiv{Department of Physics and Astronomy}, \orgaddress{Gower Street, London, WC1E 6BT, UK}}
\address[2]{\orgname{Max Planck Institute for Radio Astronomy},  \orgaddress{Auf dem H\"ugel 69, 53121 Bonn, Germany}}

\articletag{Chapter Article tagline: update of previous edition,, reprint..}

\maketitle

\begin{glossary}[Glossary]

\term{Baryon cycle:} The ``life-cycle" of gas in galaxies, from accretion into the ISM from the outside environment, through processing in molecular clouds and star formation, and finally the return of some of this gas to the external environment through outflows.

\term{Circumgalactic medium:} The mostly hot and ionised gas that fills the space outside of the stellar component of the galaxy, but still within its virial radius.

\term{Cosmic noon:} The period in the history of the universe where the star formation activity was at its highest. It corresponds to a redshift $z\sim2$, or look-back time of $\sim10$ billion years. 

\term{Cosmic web:} The interconnected filamentary structure of dark matter and gas that fills the Universe.  Galaxies form in the high density regions within the cosmic web. 

\term{Early-type galaxy:} Morphological classification referring to elliptical galaxies. 

\term{Intergalactic medium:} The extended, diffuse and hot gas component that fills the space between galaxies, tracing the underlying dark matter cosmic web. 

\term{Interstellar medium:} The multi-phase, dusty gaseous material that fills the space between the stars inside galaxies.

\term{Late-type galaxy:} Morphological classification referring to spiral and irregular galaxies. 

\term{Mass-loading factor:} The ratio between the rate of mass loss through gas outflows and the star formation rate of a galaxy, giving an indication of the potential of an outflow having a significant impact on future star formation activity through removal of ISM gas. 

\term{Quenching:} The process of stopping star formation activity in a galaxy, through either the removal or rapid consumption of its cold ISM gas, the halting of further gas accretion, or the suppression of star formation out of any available cold gas. 

\term{Quiescent galaxy:} A galaxy with very low levels of star formation activity. 

\term{Star formation main sequence:} In the context of galaxy population studies, the ``main sequence" refers to the relation between the stellar mass and star formation rate of galaxies. ``Main sequence galaxies" is commonly used as way to refer to steadily star-forming galaxies as opposed to starbursting galaxies, or those with very little star formation activity.

\end{glossary}

\begin{glossary}[Nomenclature]
\begin{tabular}{@{}lp{34pc}@{}}
ALMA & Atacama Large Millimeter/submillimeter Array\\
CGM &Circumgalactic medium\\
CNM & Cold neutral medium \\
ETG & Early-type galaxy \\
FIR & Far-infrared \\
GMC & Giant molecular cloud\\
HIM & Hot ionised medium \\
IGM & Intergalactic medium\\
ISM &Interstellar medium\\
JWST & James Webb Space Telescope \\
LTG & Late-type galaxy\\
MIR & Mid-infrared\\
PAH & Polycylic aromatic hydrocarbon\\
SFE & Star formation efficiency\\
SFR & Star formation rate\\
SKA & Square Kilometre Array \\
SNe & Supernova explosion\\
WIM & Warm ionised medium\\
WNM & Warm neutral medium\\
\end{tabular}
\end{glossary}

\begin{abstract}[Abstract]
The interstellar medium (ISM) is the material that fills the space between the stars in all galaxies; it is a multi-phase medium in pressure equilibrium, with densities and temperatures covering over 6 orders of magnitude. Although accounting for only a small fraction of the mass of any given galaxy, it is a vital component, since it holds the material responsible for galaxy growth through star formation. Studying the ISM requires careful observations at all wavelengths of the electromagnetic spectrum. This article describes the multi-phase nature of the ISM, and then puts it in the context of galaxy evolution models, emphasizing the importance of the cycling of baryons in and out of galaxies. Within this framework, the ISM plays a central role: it connects the physical processes operating on very large physical- and time-scales which control the accretion of gas onto galaxies, and the small scale processes that regulate star formation. 
\end{abstract}

\begin{BoxTypeA}[chap1:box1]{\large Key points}
\begin{itemize}
\item The {\bf interstellar medium (ISM)} is the material that fills the space between stars within galaxies.  It is typically made of 99\% gas, and $<1\%$ dust particles. 
\item The ISM is a {\bf multi-phase medium}, comprised of specific components with characteristic temperatures ranging from 10 to $10^6$~K. 
\item Studying this multi-phase medium requires {\bf observations across the electromagnetic spectrum}.  Gaining a full picture of the ISM in galaxies is therefore a technical challenge.  
\item The {\bf cold neutral medium} has a characteristic temperature of 10-100~K and the highest density within the ISM, and is the birth place of new stars. 
\item The {\bf warm medium} has typical temperatures of 5000-10000~K and has both an atomic and an ionised component. 
\item The {\bf hot ionised medium} is the component with typically fills a large fraction of the volume, but only accounts for a small fraction of the mass of the ISM, due to the low densities and high temperatures reached through heating by supernova explosions. 
\item The relative contribution of the different phases to the overall ISM mass budget varies from galaxy to- galaxy, with {\bf spiral and elliptical galaxies} forming two markedly different populations in terms of their ISM. 
\item {\bf Dwarf galaxies} typically have more mass in their ISM than the amount of mass they have in stars. 
\item Observing the ISM in {\bf high-redshift galaxies} (i.e. galaxies we are seeing as they were only hundreds of millions of years after the Big Bang) is a challenge, but an area of intense study especially with the {\bf James Webb Space Telescope (JWST)}. 
\item The {\bf baryon cycle} refers to the lifecycle of gas which flows into galaxies along the cosmic web, cools and settles into the ISM where it can contribute to star formation,  and finally partly returned to the environment through energetic winds powered by young stars and supermassive black holes. 
\item The ISM provides astronomers with vital information regarding all components of the baryon cycle.  Put together, this evidence supports a {\bf model for galaxy evolution} where the baryon cycle plays a crucial role. 
\end{itemize}
\end{BoxTypeA}

\section{Introduction}
\label{sec:intro}

The interstellar medium (ISM) is defined as the dusty, gaseous material that permeates the space between the stars within galaxies. Judged purely in terms of its mass, the ISM might seem insignificant: it accounts for only $\sim 10-20\%$ of the total baryonic mass (i.e. the ``normal" matter) of a typical star-forming galaxy in the Universe today, and as little as $\sim 1\%$ in the case of a typical elliptical galaxy. The numbers shrink significantly further if we consider the bigger picture, with the ISM making up only about $\sim 1.5\%$ of the overall baryonic matter budget of the Universe today \citep{saintonge22}. Even though the ISM constitutes only a small fraction of a galaxy's mass, it plays a central role in regulating the formation and growth of stars and galaxies. The ISM is also a rich trove of information that allows us to reconstruct the star formation and chemical enrichment histories of galaxies. In this chapter, after providing a brief overview of the composition and structure of the ISM, we describe its central role within our galaxy evolution framework.

\section{Composition and structure of the interstellar medium}
\label{sec:overview}

The ISM of a typical star-forming galaxy, like the Milky Way, spans a very wide range of physical conditions, with temperature and density varying by at least six orders of magnitude. This mere fact implies that the ISM is not in thermal equilibrium. Instead, it is pressure equilibrium that regulates the ISM, leading to its break down into three distinct phases \citep{mckee77}: the hot ionised medium, the warm medium (including both a neutral and an ionised component), and the cold neutral medium out of which cold, dense molecular clouds also form. Table \ref{tab:MW} gives an overview of the characteristic temperatures and densities of these five components in a Milky Way-like galaxy in the Universe today. 

\begin{table}[h]
\TBL{\caption{Characteristic properties of the phases of the ISM in a Milky Way-like galaxy at the present time}\label{tab:MW}}
{\begin{tabular*}{\textwidth}{l l l l l}
\toprule
\multicolumn{1}{l}{\TCH{Phase}} &
\multicolumn{1}{l}{\TCH{T [K]}} &
\multicolumn{1}{l}{\TCH{n [$cm^{-3}$]}} & 
\multicolumn{1}{l}{\TCH{Mass fraction}} & 
\multicolumn{1}{l}{\TCH{Typical observational tracers}} \\
\colrule
Hot ionised medium 		& 1,000,000  	&  0.004  	& 3\%  	&  X-ray emission, UV absorption lines  \\
Warm ionised medium 	& 8,000  		&  0.2  	& 12\%  	& Optical emission lines, non-thermal radio emission \\
Warm neutral medium 	& 6,000 		& 0.4  	& 35\%  	& \hi~21~cm line \\
Cold neutral medium 	& 80  		& 40  	& 30\%  	& \hi~21~cm line, UV/optical absorption lines, FIR fine structure lines \\ 
Molecular clouds 		& 15  		& $>$100 	& 20\% 	& Molecular lines (CO in particular), FIR continuum (dust emission) \\
\botrule
\end{tabular*}}{
\begin{tablenotes}
%\footnotetext[a]{Table footnote text...}
\footnotetext{\source{\citet{tielens05, ryden21}}}
\end{tablenotes}
}
\end{table}

\begin{figure}[h]
\centering
\includegraphics[width=1.0\textwidth]{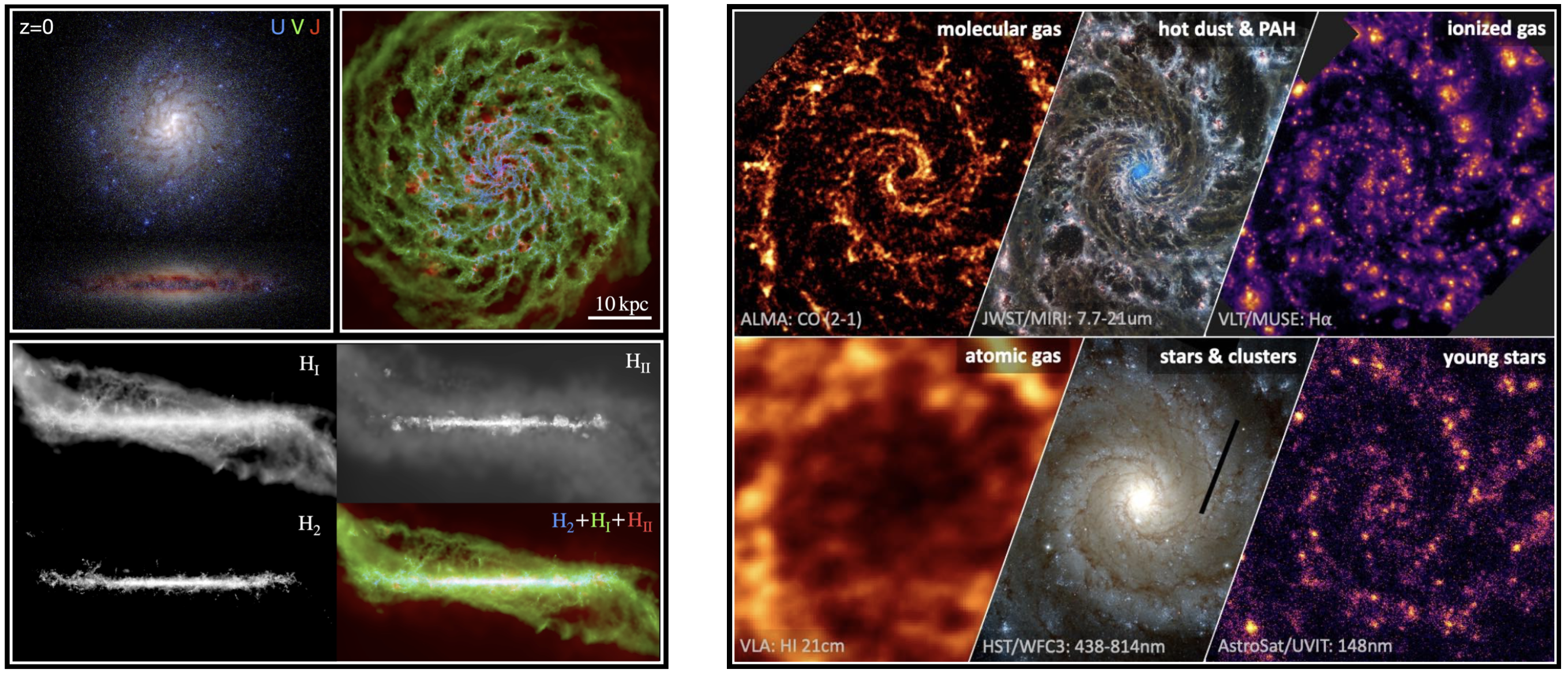}
\caption{{\bf Left:} Visualisation of the multi-phase structure of the ISM in a Milky Way-like galaxy at $z=0$ in the FIREbox simulation \citep{hopkins18}. {\it Top left:} Face-on and edge-on view of the stellar and dust components. {\it Top right:} Color composite image of the galaxy's molecular (\hmol, blue), atomic (\hi, green) and ionised (\hii, red) hydrogen. {\it Bottom:} Edge-on view of the three different ISM components. Figure reproduced from \citet{feldmann23}. {\bf Right:} Observations of the multi-phase ISM and stellar population in the nearby galaxy NGC628. Image credit: PHANGS Team / J. Sun (Princeton) / J. Schmidt (ASCL) / NASA / ESA / CSA. }
\label{fig:ISMstructure}
\end{figure}

\subsection{The cold interstellar medium}
\label{sec:coldISM}

The coldest phases of the ISM, the cold neutral medium (CNM) and the molecular clouds, dominate the total ISM mass budget. The CNM is distributed in sheets and filaments, which are the birthplace of the giant molecular clouds (GMCs) in which stars eventually form. The filamentary structure of the neutral medium, with molecular clouds embedded, is illustrated in Figure \ref{fig:ISMstructure}. 

Given the range of temperatures and densities spanned by the cold ISM, a number of different tracers need to be used. Given that the ISM is $\sim70\%$ hydrogen by mass, tracers that directly detect the hydrogen emission are optimal. In the CNM, hydrogen is in its atomic form (H{\sc i}), and therefore is directly observable at radio wavelengths through the 21~cm line, either in emission or absorption. A challenge is that the warm neutral medium (WNM) also is dominated by atomic hydrogen emitting the 21~cm line; this cold and warm neutral gas can be considered a two-phase medium in pressure equilibrium \citep{wolfire03}. In the Milky Way, \hi\ observations in absorption are particularly useful to disentangle the CNM from the WNM. However, in any other galaxy of the nearby universe where \hi\ is observed in emission and (in most cases) spatially unresolved, the two phases are observed together. The CNM can also be probed through observations of absorption lines at UV and optical wavelengths against the background of a star (for the Milky Way) or quasar (for other galaxies). 

Molecular cloud structures can form through compression of the neutral medium. The lifecycle of GMCs, from formation to dissipation, is regulated by a long list of physical processes, including gravitational and dynamical interactions with other clouds and the galactic potential (e.g. spiral arms), magnetic fields, pressure and turbulence \citep{chevance20review}. Once again, most of the molecular medium by mass is hydrogen.  Because \hmol\ is symmetric, the vibrational and rotational states of the molecule can only emit through their quadrupole moment, which requires gas temperatures well above the typical temperatures of molecular clouds (or very strong UV radiation fields).  As these conditions are not readily met, this means that cold \hmol\ cannot be directly observed.  Instead, the cold molecular ISM is observed through rotational lines of other molecules such as CO,  HCN, HCO$^+$, CS, and H$_2$O. In fact, more than 200 different molecules have been detected in the ISM of the Milky Way, but only a small subset of those are typically bright enough to be detectable in extragalactic objects.  As the second most abundant molecule in the ISM, the CO molecule is the workhorse of extragalactic molecular gas studies. With an excitation temperature of 5.5~K and a critical density of $\sim3000$~cm$^{-3}$, the $^{12}$CO($J=1\rightarrow0$) transition is particularly well suited to probe the bulk of the molecular ISM (see Tab. \ref{tab:MW}). At higher redshifts, it is common practice to target higher transitions (e.g. $J=3\rightarrow2$ at $z\sim2-3$), as these lines are redshifted into convenient atmospheric windows.  Observations and simulations of Milky Way-like galaxies show that the molecular ISM is confined to a very thin disc, with the GMCs tracing the regions of highest density (Fig.~\ref{fig:ISMstructure}).

There is dust mixed in within the cold phases of the ISM, both neutral and molecular. Typically, the dust-to-gas mass ratio is of the order of $\sim1\%$, but with significant variations under specific conditions such as low metallicity or strong radiation fields.  Dust emits thermally in the infrared; hot/warm dust grains heated by star formation or active galactic nuclei (AGN) activity emit most strongly in the mid-infrared (MIR), and cold dust with characteristic temperatures of 10-20~K (or up to $\sim50$~K at higher redshifts) peaks in the far-infrared (FIR).  A specific family of very small carbon grains, the polycyclic aromatic hydrocarbons (PAHs), have strong spectral emission and absorption features in the MIR. Through these PAH features, the James Webb Space Telescope (JWST) is able to map dust emission in nearby galaxies (see example in Fig. \ref{fig:ISMstructure}), providing an avenue of obtaining such high spatial resolution images of dust in a wide range of environments.

\subsection{The warm interstellar medium}
\label{sec:WIM}

The warm interstellar medium refers to material with typical temperatures of $\sim5,000-10,000$~K, and is composed of a neutral (the WNM, introduced briefly above) and an ionised component (the warm ionised medium, WIM).  The WNM, being not only warmer but also more diffuse than the CNM, is best observed through the \hi\ 21~cm line in emission.  Such observations (and simulations) reveal that the WNM is far more extended, and in a thicker configuration than the cooler phases of the ISM (see Fig. \ref{fig:ISMstructure}).  In many nearby spiral galaxies, including the Milky Way, these extended warm \hi\ discs are seen to warp at large radii due to a lack of support from a stellar disc.  Atomic gas distributions can indeed be seen to extend far beyond the stellar discs of the galaxies, and therefore this ISM phase can be seen as an extended reservoir of material for future star formation if it can, under the right conditions, be brought into the higher density regions of the galaxy.  

As its name implies, the WIM is dominated by hydrogen in its ionised state, due to photoionisation by hot stars.  There is a distinction between the WIM and other ionised regions of the ISM, such as \hii\ regions (the ionised Str\"omgren spheres around hot young stars) and planetary nebulae (ionised regions around young white dwarf stars).  While having similar temperatures, \hii\ regions and planetary nebulae have significantly higher densities and are confined to the thin disc, whereas the WIM is a thick ionised gas layer that permeates the discs and halos of galaxies. These various ionised gas components and their distinct spatial distributions can clearly be seen in the simulated galaxy shown in Fig. \ref{fig:ISMstructure}, and more generally they appear as an extended diffuse component in Balmer line observations of the Milky Way and nearby galaxies \citep{haffner09}.

\subsection{The hot interstellar medium}
\label{sec:HIM}

The hot ionised medium (HIM) is only a very small fraction of the total ISM mass of a Milky Way-like galaxy, but typically fills $\sim50\%$ of its volume due to its very low density. In contrast, most of the ISM mass of early-type galaxies is in the HIM (more on this in the next section).  The existence of the HIM was predicted by \citet{spitzer56}, based on the argument that in the absence of a hot gas corona providing pressure support, the \hi\ clouds that were known to exist far above the plane of the Milky Way would be over-pressurised and rapidly expand to reach equilibrium with the low density surrounding gas.  The HIM was later observed directly as diffuse X-ray emission, and through absorption lines at UV wavelengths of highly ionised metal lines (such as O{\sc vi} and C{\sc iv}) against background sources (stars or quasars). 

Based on the number of young hot stars in a typical galaxy, the expectation is that there are not enough UV photons for the high temperature and ionisation state of the HIM to be explained through photoionisation. Instead, the ionisation and high temperatures, in excess of a million K, are reached and maintained through collisions with free electrons in shocks produced by supernova explosions (SNe).  Since star formation occurs primarily in the thin disc of spiral galaxies, the colder ISM is riddled with bubbles of HIM gas, as a result of localised SNe; these bubbles can be clearly seen, delineated by the cooler ISM phases, in both the simulated and observed galaxy shown in Fig.  \ref{fig:ISMstructure}.  As a matter of fact, our own Solar System is currently passing through a HIM bubble (known as the ``Local bubble") at this point in its journey around the Milky Way.   If several SNe occur in close succession in the same region of space, their individual ionised gas bubbles can coalesce, leading to the formation of chimneys that break through the denser, cooler surrounding gas, allowing the HIM to ``spill out" into the halo.  The future fate of this gas is discussed further in Sec. \ref{sec:CGMconnection}.

\section{Census of the ISM across galaxy populations}

Due to the wide range of temperatures and densities, observing all the components of the ISM requires data from across the entire electromagnetic spectrum (see overview in Tab. \ref{tab:MW}).  All of these observations come with specific challenges, and with particulars that affect their availability depending on the galaxy type and cosmic lookback time.  For all these reasons, we do not yet have a complete and robust census of the ISM (in terms of its overall properties and the relative breakdown in the different phases) across the whole galaxy population and with redshift.  However, based on the information available, some general trends are emerging.

\subsection{Star-forming galaxies of the nearby Universe}

Given its central role in fuelling star formation activity, the cool ISM of extragalactic systems has been the focus of extensive studies for decades.  This includes studies of the cold and warm neutral gas with \hi-21~cm observations, and of the molecular gas through mostly CO line observations. Alternatively, thermal dust emission in the FIR and sub-millimeter wavelength ranges is often used to infer the mass and distribution of the cold atomic and molecular gas, under the assumption that dust is well mixed with these ISM phases.   

Initially, observations were focused on the most star-forming and gas-rich galaxies, but with the increased sensitivity and speed of modern instruments, we now have a far more complete picture.  Of particular importance in shaping this view of the atomic and molecular ISM of nearby galaxies have been the wide-area blind HI surveys performed with the Parkes and Arecibo radio telescopes \citep[e.g.][]{HIPASS,alfalfa1}, as well as targeted surveys aimed at providing homogeneous atomic and molecular gas measurements for large optically-selected galaxy samples \citep[see overview in][]{saintonge22}. 

\begin{figure}[h]
\centering
\includegraphics[width=1.0\textwidth]{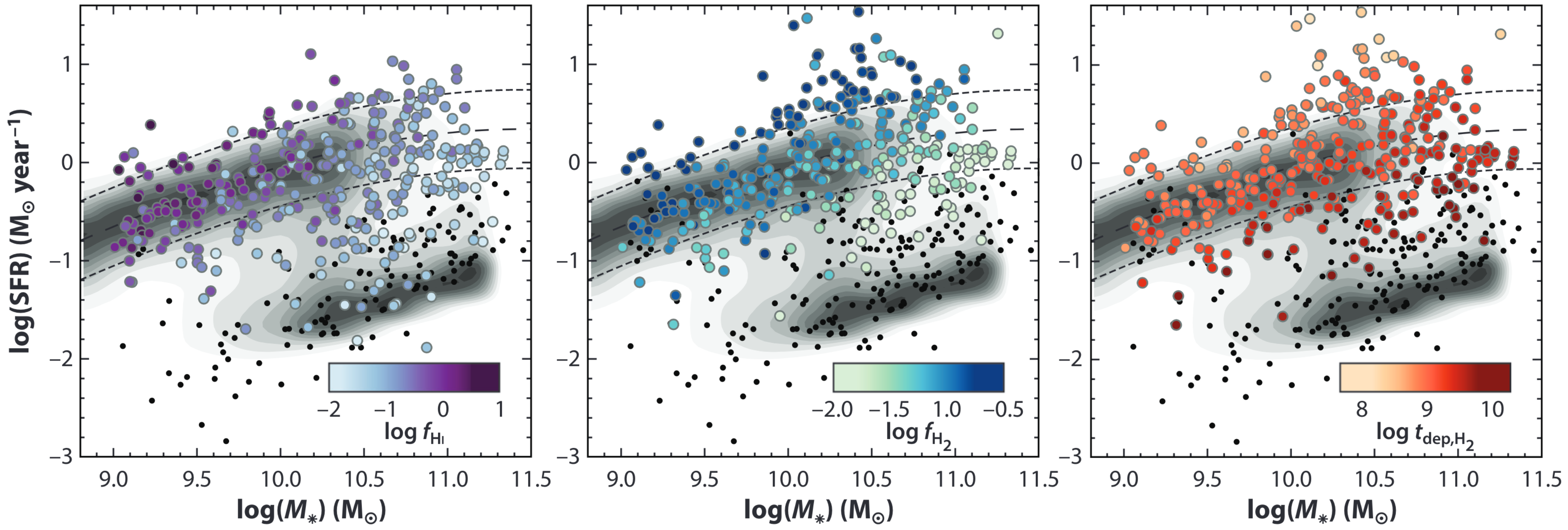}
\caption{Atomic and molecular gas across the local galaxy population.  In each panel, the galaxies of the xCOLDGASS survey \citep{saintonge17} are shown in comparison to the full galaxy population at low redshift (SDSS; gray countours), and color-coded by a specific ISM-related parameter (black dots are non-detections in the H{\sc i} and/or CO emission lines). The quantities shown are the atomic gas mass fraction (\fhi$\equiv$\mhi/\mstar; left), the molecular gas mass fraction (\fgas$\equiv$\mh/\mstar; middle), and the molecular gas depletion timescale (\tdep$\equiv$\mh/SFR; right). The solid and dashed lines shows the position and width of the main sequence. Figure reproduced from \citet{saintonge22}.}
\label{fig:gasfractions}
\end{figure}

Overall, the picture that emerges from these observations is relatively simple and unsurprising: the more gas-rich the galaxy is, the higher its star formation rate.  However, when looking in more detail, the picture is more subtle, revealing a series of triggers and bottlenecks determining the balance between the atomic and molecular phases of the ISM, and the rate of star formation.  This is partly illustrated in Fig. \ref{fig:gasfractions}.   

Star-forming galaxies trace a tight relation in the 2D plane formed by stellar mass and star formation, known as the ``main sequence of star formation" (dashed line in Fig. \ref{fig:gasfractions}).  In terms of morphology, the vast majority of main sequence star-forming galaxies are spirals (also referred to as ``late-type galaxies" or LTGs).  Along the main sequence, galaxies with lower stellar mass (\mstar$\sim10^9$\msun) typically have comparable masses of HI gas (\mhi)  (i.e. their atomic gas fraction, \fhi$\equiv$\mhi/\mstar, is typically $\geq 1$, see Fig. \ref{fig:gasfractions} left panel). By the time main-sequence galaxies reach a mass of $\sim5\times10^{10}$\msun\ (which is roughly the mass of the Milky Way, as a reference point), the atomic gas fraction drops to only about 10\%.  In contrast to this, the molecular gas fraction (\fgas$\equiv$\mh/\mstar; middle panel of Fig. \ref{fig:gasfractions}) is much more constant along the main sequence, with an average value of 10\% across most of the mass range probed by current surveys, with a drop observed only at the highest masses in a way that naturally explains the flattening of the main sequence.  The gradual change in the balance between atomic and molecular gas masses along the main sequence is caused by two main factors: (1) a reduction in the rate of gas accretion in galaxies with stellar masses greater than $\sim 10^{10}$\msun\ manifesting itself as a drop in \fhi\ towards the massive end of the main sequence, and (2) a change in the incidence of factors such as spiral arms, bars and bulges, which affect the dynamics and pressure conditions in the ISM and therefore the atomic-to-molecular conversion. The right panel of Fig. \ref{fig:gasfractions} is discussed in Sec. \ref{sec:SFconnection}.

\subsection{Quiescent galaxies of the nearby Universe}

Any galaxy with a star formation rate (SFR) less than approximately a third of the value of a main-sequence galaxy of this given mass is generally considered to be ``quiescent", in the sense that it has a low rate of star formation. One often differentiates between galaxies just below the main sequence and probably having recently experienced a drop in star-forming activity (generally referred to as recently-quenched or ``green valley" galaxies), and those which have at least an order of magnitude less star formation than a main-sequence galaxy of the same mass and which are considered to be ``quenched", referring to the very low levels of star formation activity, and low cold ISM content.  In terms of morphology, the vast majority of these quiescent galaxies are lenticulars and ellipticals, a class of galaxies commonly known as ``early-type" galaxies (ETGs).  Given the tight link between the atomic and molecular gas phases of the ISM and star formation activity, it generally goes hand in hand that quiescent galaxies are also poor in WNM and CNM gas.  Instead, most of the ISM of quiescent galaxies is in the HIM phase.   

The processes that can turn a star-forming galaxy into a quiescent one are collectively referred to as ``quenching mechanisms".  These mechanisms all work by suppressing star formation in three main ways: heating and/or ejecting the cold ISM from galaxies (via stellar and AGN feedback), preventing the replenishment of the WNM and CNM by halting cold gas accretion, or stabilising any cold atomic and molecular gas from fragmenting and collapsing into stars.  Studying the distribution and kinematics of the various ISM phases can shed light on the specific quenching mechanisms that have affected any given quiescent galaxy.

\begin{figure}[h]
\centering
\includegraphics[width=1.0\textwidth]{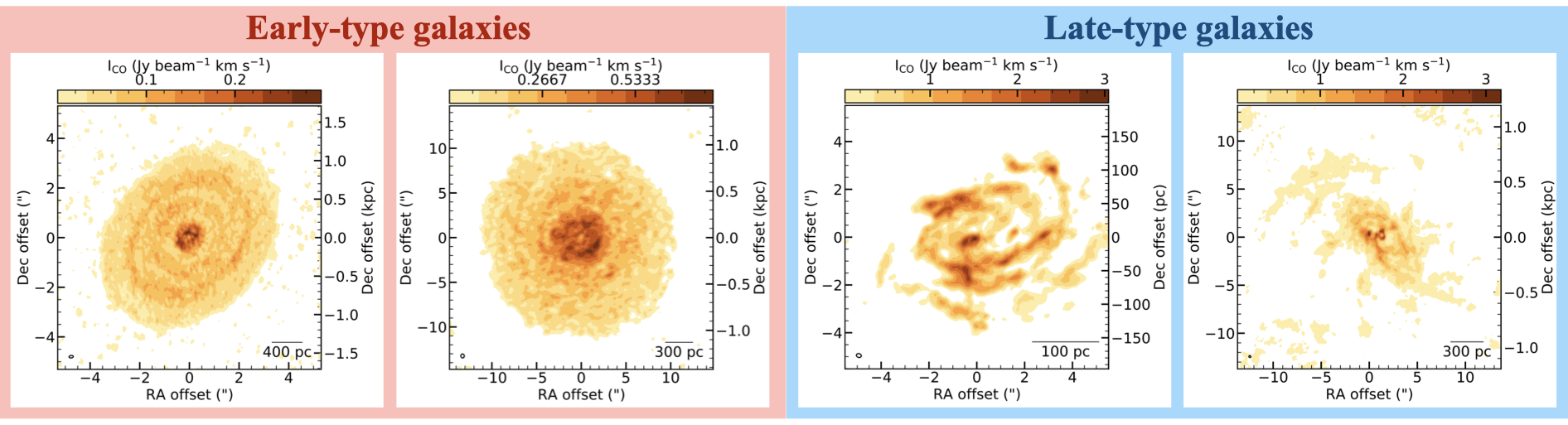}
\caption{Molecular gas in the central regions of two nearby ETGs (left) and LTGs (right), as imaged with ALMA.  The striking differences in terms of clumpiness highlight the fact that low SFRs in ETGs are not due simply to the absence of molecular gas, but also to the stability of this gas against fragmentation and star formation. Figure adapted from \citet{davis22}.}
\label{fig:COinETGs}
\end{figure}

As can be seen in Figure \ref{fig:gasfractions}, some quiescent galaxies can have atomic gas-to-stellar mass ratios as high as 10\%.  On average, even the most massive, least star-forming elliptical galaxies have an atomic gas mass ratio on the order of $\sim 1$\%, showing that their very low star formation activity is not due entirely to the absence of WNM and CNM gas, but also to the low star formation efficiency out of this gas.  This is illustrated in Figure~\ref{fig:COinETGs}, which shows the molecular gas in the central region of four nearby galaxies: two ETGs and two LTGs. The smoothness of the molecular gas discs in ETGs illustrates the bottleneck to star formation in these types of objects: the stability of the molecular discs because of the massive stellar bulge, preventing them from fragmenting and triggering star formation.  This process of lowering the star formation efficiency through the action of massive stellar bulges preventing fragmentation and collapse of the gas is generally referred to as ``morphological quenching" \citep{martig09}.

\subsection{Dwarf galaxies of the nearby Universe}

The exact definition of a ``dwarf galaxy" is subjective, but generally galaxies with masses similar to (and smaller than) that of the Large Magellanic Cloud (a few $\times10^9$\msun) are considered to fall into this category.  Like their more massive counterparts, they come in both the star-forming and quiescent varieties, although the fraction of quiescent galaxies is significantly lower in this mass range, with quenching mostly the result of interactions of galaxies with their large-scale environment \citep{cortese21}.  Star-forming dwarf galaxies (morphologically classified as Dwarf Irregulars, or ``dIrr"s) are in fact the most common type of galaxies in the Universe.  Quiescent dwarf galaxies (traditionally described morphologically as Dwarf Spheroidals, dSph, or Dwarf Ellipticals, dE) are not discussed here, given that their ISM is very sub-dominant. 

In terms of their ISM, dwarf galaxies are remarkable compared to more massive systems in having their baryonic mass budgets dominated by the ISM, with the total mass of H{\sc i}-emitting gas surpassing even the stellar mass.  This gas tends to be distributed in extended, thick discs.  Because of their low metal content, dwarf galaxies have lower dust contents compared to more massive galaxies (i.e. lower dust-to-gas mass ratios), and in the absence of dust to provide shielding from the harsh UV radiation field, the CO molecules are easily photo-dissociated.  The resulting difficulty in measuring both thermal dust continuum emission in the FIR and CO emission lines makes the study of the CNM and molecular clouds in dwarf galaxies an infamously difficult problem to solve. As such, the question of whether the relatively low star formation rates in these galaxies compared to the total mass of gas as measured from HI is due to a bottleneck in conversion of material from the WNM to the CNM (and into the denser molecular clouds), or due to a lower star formation efficiency out of any molecular gas present, remains open and debated \citep[a recent review of this topic is presented in][]{hunter24}.

\subsection{Galaxies beyond the nearby Universe}
\label{sec:highz}

Only in the low redshift Universe do we have at the moment the ability to directly observe all of the ISM phases. The best-studied ISM phases beyond the local Universe are the dense molecular gas, either from observations of CO emission lines or through thermal dust emission, and the WIM observed through in particular rest-frame optical emission lines.  These emission lines are red-shifted into the near-infrared for galaxies at $z\sim2$ (an epoch referred to as ``cosmic noon", see next paragraph), and into the MIR at even higher redshifts; JWST, with its near- and mid-infrared instruments, is revolutionizing our ability to study the warm ISM in galaxies of the very early Universe. 

\begin{figure}[h]
\centering
\includegraphics[width=1.0\textwidth]{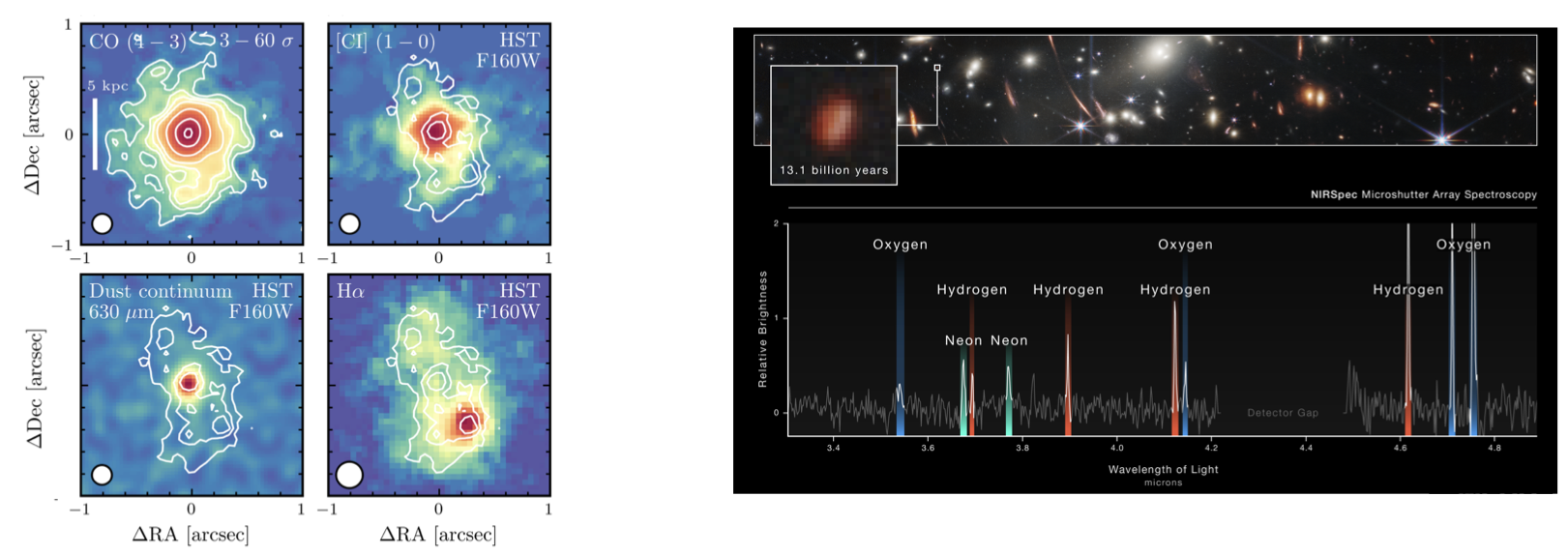}
\caption{Observing the ISM of high redshift galaxies. {\bf Left:}  Multiple phases of the ISM in BX610, a massive galaxy at cosmic noon ($z = 2.21$), including the molecular ISM (CO(4-3)), the neutral medium ([C{\sc i}](1-0)), the ionised medium (H$\alpha$) as well as dust continuum. Figure reproduced from \citet{arriagadaneira24}. {\bf Right:} JWST NIRSpec spectrum of a $z=8.5$ galaxy, highlighting the chemical complexity of the ISM only 600 million years after the Big Bang.  Image credit: NASA, ESA, CSA, STScI. }
\label{fig:highz}
\end{figure}

Star formation activity in the Universe reached its peak at a redshift of $\sim2$, corresponding to a look-back time of 10~Gyr. At this epoch, the star formation rate per unit volume is at its highest, with the star formation rate in individual galaxies elevated by a factor of $\sim 10$ compared to local galaxies \citep[e.g.][]{madaudickinson14}.  Over the years, different explanations have been put forward to explain this rise in star formation activity at earlier times, often invoking galaxy mergers, but the debate has been convincingly settled by molecular gas and dust observations showing a comparable increase in the mass of gas available for star formation \citep[see review in ][]{tacconi20}.  Star-forming main sequence galaxies at $z\sim2$ are typically rotation-dominated discs, just like at lower redshifts, but with significantly higher gas fractions (\fgas\ higher by a factor of 5-10 compared to $z=0$) and slightly elevated star formation efficiency \citep{forster20}.  Figure \ref{fig:highz}~(left) shows maps of the molecular, atomic and ionised ISM in a characteristic massive star-forming galaxy at cosmic noon. 

At even higher redshifts ($z>3-4$), the cold ISM can be probed still through CO emission line observations, although the declining metals and dust contents of galaxies at these earlier epochs makes the detectability of these lines a challenge.  Another possible avenue at these redshifts is to use FIR fine structure lines, which are redshifted into some accessible atmospheric bands.  Of particular interest is the ionised carbon line ([C{\sc ii}]) at a rest wavelength of 158$\mu$m; being the dominant cooling line of cool interstellar gas, it is bright and mostly unaffected by dust attenuation.  Oxygen lines in the FIR are also commonly used (e.g. the [O{\sc iii}] line at 88$\mu$m and the [O{\sc i}] line at 63$\mu$m), all showing good correlation with the SFR.  Recent JWST observations of the WIM through rest frame optical lines, and follow-up observations of FIR lines with facilities such as the Atacama Large Millimeter/submillimeter array (ALMA), are providing us with a view on the ISM of galaxies at very high redshifts ($z>6-8$) like never before (see example spectrum in Fig. \ref{fig:highz}, right panel).  The initial results suggest that the ISM is already significantly metal-enriched only a few hundred million years after the Big Bang, challenging current models for star formation and dust production.

\section{The ISM as a component of the baryon cycle}
\label{sec:baryoncycle}

\begin{figure}[h]
\centering
\includegraphics[width=1.0\textwidth]{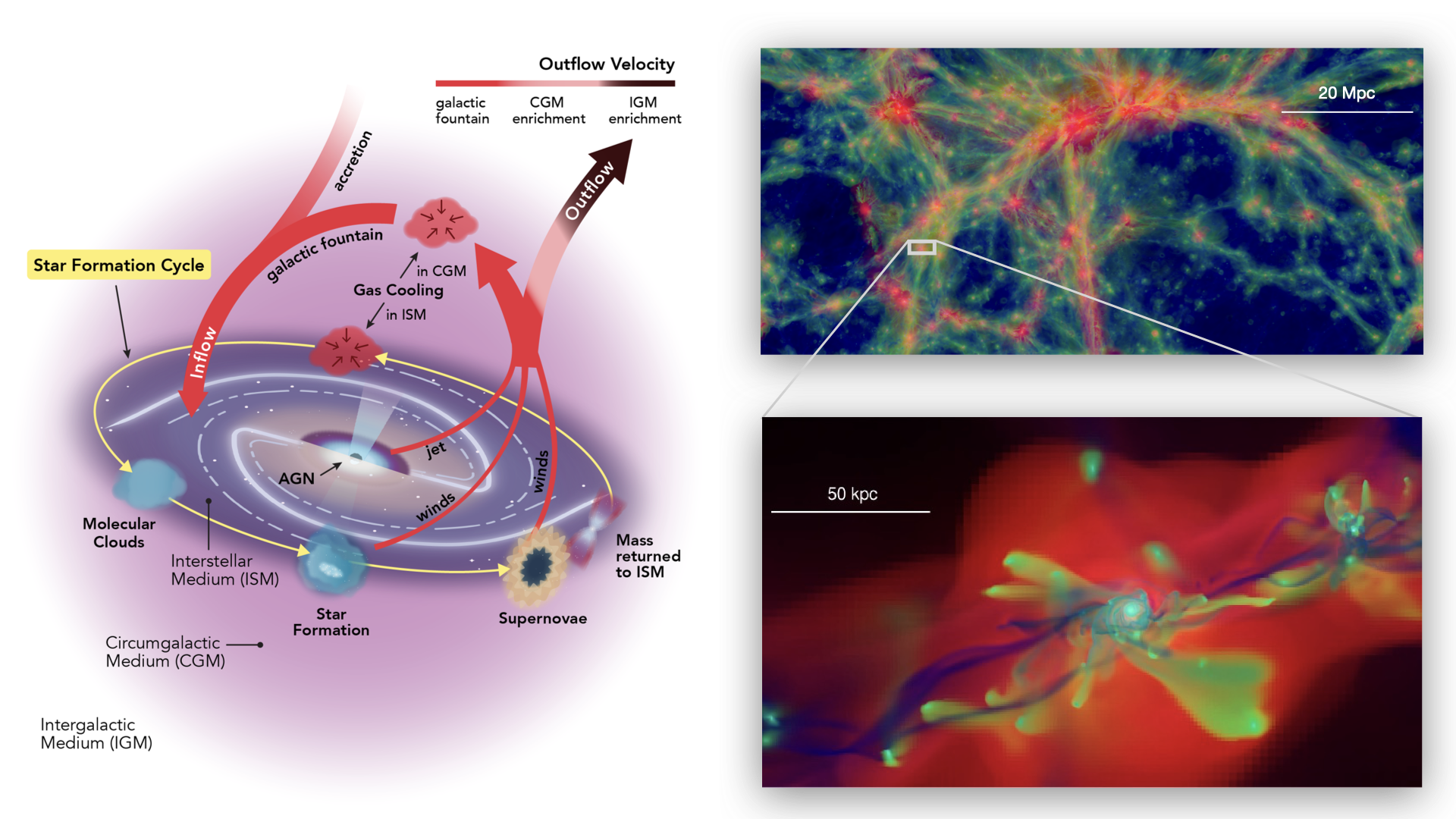}
\caption{{\bf Left:} A representation of the baryon cycle, highlighting gas inflow, the processing of some of this gas into stars within the ISM, and the stellar- and AGN-powered outflows. Image credit: L.Davies(ICRAR)/M.Romano(MPIfR)/M.Saraf(UCL)/A.Vadolia/NASA.  {\bf Right:} Examples of numerical simulations showing the large scale gaseous environments around galaxies. {\it Top:} On very large scales, the IGM gas follows the cosmic web, with clear temperature structure (blue: cool, green: warm, red: hot). {\it Bottom:} A different, zoomed-in simulation, showing the streams of cold, metal-poor gas (blue), the hot CGM gas surrounding the galaxy (red), and the metal-rich gas being either stripped from smaller satellite galaxies, or expelled through outflows (green).  Figures from \citet{schaye15} and \citet{agertz09}.}
\label{fig:baryoncycle}
\end{figure}

Within the field of galaxy evolution, more and more emphasis has been given to studies of the ISM, which is resulting in a corresponding evolution of the models and of the framework used to put these observations into context. This change of perspective has been driven by increased observational capabilities, as well as theoretical and numerical efforts.  There is a family of theoretical models, generally referred to ``equilibrium models" for galaxy evolution, which have enjoyed significant success in explaining several galaxy scaling relations, their evolution, and their interconnections by invoking only a small number of physical processes \citep[e.g.][]{lilly13}.  In these models, galaxies are thought of as gas reservoirs that follow simple mass conservation principles; the accretion rate of new gas onto the reservoir must be balanced out by the rate at which gas is converted into stars and the rate at which gas is ejected from the reservoir through star formation-driven outflows.   This process of gas accretion from the intergalactic medium (IGM; the hot and diffuse gas that fills the space in between galaxies), through the CGM and into the ISM, its subsequent participation in the star formation process, and the return of material to the CGM (or even the IGM) through outflows is generally referred to as the ``baryon cycle" and is a subject of intense research activity. An illustration of the baryon cycle and of its main components is presented in Figure \ref{fig:baryoncycle}~(left). A key challenge is that the inflow and outflow components of the baryon cycle are notoriously difficult to observe.  The ISM is therefore of particular importance: from a physical point of view, it sits at the intersection of the large range of physical scales that drive the baryon cycle and directly enables star formation, and from an observational point of view, the ISM encodes significant information about all aspects of the baryon cycle, which can be retrieved through measurements of the mass and composition of its various phases.

\subsection{Connection to the intergalactic and circumgalactic medium}
\label{sec:CGMconnection}

Observations show that the characteristic depletion timescales for the cold ISM are on the order of a few billion years, which is significantly shorter than the age of the Universe.  This means that assuming a constant SFR, galaxies would consume all their ``fuel" within a few billion years.  The fact that galaxies continue forming stars over significantly longer periods, and to the present day, implies that their gas reservoirs must be steadily resupplied \citep{tacconi20}.  Numerical hydrodynamical simulations of galaxy formation provide important clues into how this replenishment of the ISM proceeds (see examples of such simulations in Fig. \ref{fig:baryoncycle}, right panel). The accretion of gas onto galaxies is generally referred to as either ``smooth" or ``clumpy", with the former referring to gas being accreted smoothly, over long timescales, and the latter referring to gas acquired through the accretion of low mass satellite galaxies.  Smooth accretion is the dominant factor contributing to the mass assembly history of galaxies, with mergers however playing an important role, especially for massive galaxies and at high redshifts. Simulations further reveal a drop in the efficiency of this smooth accretion in galaxies with dark matter halo masses $\gtrsim 10^{12}$\msun\ due to the heating of the incoming gas as it encounters the extended warm CGM of these galaxies \citep[e.g.][]{keres05}.  

Another source of accreting gas is material ejected from galaxies through outflows, which subsequently falls back onto the disc.  This phenomenon is generally referred to as ``galactic fountains" (see Fig. \ref{fig:baryoncycle}., left)  As already described in Sec.~\ref{sec:HIM}, the combined action of SNe can lead to the formation of chimneys which can break through the denser, cooler surrounding gas, ultimately spilling out into the hot CGM gas.  Theory suggests that the ejected gas can trigger the cooling of some of the CGM gas (``condensation''), which then falls onto the disc where it is accreted into the ISM \citep{fraternali17}.  

It is challenging to test these ideas and identify the origins of accreting gas with direct observations of the accreting gas flows, evidence (albeit more indirect) can also be found in observations of the interstellar medium.  For example, gas coming from the IGM through smooth accretion and material accreted through a galactic fountain process are likely to have distinct chemical compositions, with the former generally taken to be relatively metal-poor, and the latter being already pre-enriched to a higher degree.  Observations of the abundance ratio and overall metal content of the ISM are therefore a way to gain indirect evidence into the nature of the mechanisms refuelling the ISM under different conditions.  Gas accreted through different channels also bears different kinematic signatures.  Another avenue to better understand ISM accretion is to combine studies of the CGM in-situ \citep[typically through observations of Hydrogen Ly$\alpha$ or ionised metals such as carbon and magnesium along the line of sight of quasars, see e.g.][]{tumlinson17} with measurements of the mass and extent of the molecular and neutral ISM, to directly explore their connection \citep[e.g.][]{borthakur24}.  Such observations, as well at the opportunities offered by the Square Kilometer Array (SKA) to measure the \hi\ 21~cm emission at higher redshifts than currently achievable, will be crucial over the next decade to help us improve our understanding of exactly how the ISM is replenished in galaxies of different masses and at different cosmic times. 

\subsection{Connection to star formation and feedback}
\label{sec:SFconnection}

There is a very tight connection between the ISM and the star formation process: not only does the ISM provide the raw material for star formation, it also sets the conditions required for cloud formation and collapse.  In return, the metals and energy produced through various stages of the star formation and stellar evolution processes shape the surrounding ISM.  

The process of converting gas into stars is often quantified through its efficiency.  Care must however be taken to avoid confusion between two distinct definitions of ``star formation efficiency" (SFE) commonly used across the literature.   The first definition, typically used when discussing galaxy-scale measurements, relates the current star formation rate to the total molecular ISM mass: 
\begin{equation}
{\rm SFE}_{H_2} \equiv t_{dep, H_2}^{-1} = {\rm SFR} / M_{H_2}
\end{equation}
The second definition is mostly used when talking about the star formation process at molecular cloud scales.  Given that the natural comparison timescale for a collapsing cloud is the free-fall time, it is common to define the star formation efficiency as the fraction of the gas in a molecular cloud converted to stars per unit free-fall time ($\tau_{ff}$): 
\begin{equation}
\epsilon_{ff} \equiv (SFR / M_{H_2})~\tau_{ff}
\end{equation}

This cloud-scale star formation efficiency is commonly estimated to be $\epsilon_{ff} \sim 1-2\%$.  Given that molecular clouds appear to survive for a few free fall times before being dispersed by their young stars, the overall star formation efficiency of a given molecular cloud is estimated to be $\sim 5-10\%$, highlighting the inefficiency of the gas-to-stars conversion process  \citep[e.g.][]{chevance20review}.    On the scale of entire galaxies, the characteristic molecular gas depletion time is $t_{dep, H_2} \sim 1$~Gyr, corresponding to ${\rm SFE}_{H_2} \sim 10^{-9}$ yr.  As can be seen in the right panel of Fig. \ref{fig:gasfractions}, there are however some systematic deviations from this characteristic value: highly star-forming galaxies, above the main sequence, typically have shorter depletion times (i.e. higher ${\rm SFE}_{H_2}$), whereas quiescent galaxies tend to have lower than average ${\rm SFE}_{H_2}$.  As mentioned in Sec. \ref{sec:highz}, this quantity is also a slowly evolving function of redshift with ${\rm SFE}_{H_2}(z) \sim (1+z)^{1.5}$, closely following the dynamical timescale of the Universe as set by the Hubble time \citep{tacconi20}.  Understanding the mechanisms behind the systematic variations in  ${\rm SFE}_{H_2}$ and the spatial- and time-scales responsible for regulating star formation activity is a very active current area of research.  The ability to resolve molecular clouds in external galaxies (see e.g. Fig \ref{fig:ISMstructure}), therefore probing a much broader range of galactic environments (in terms of pressure, metallicity, radiation fields, dynamics, etc.) compared to what is accessible within the Milky Way, is a crucial development driving our understanding in this field. 

``Feedback" refers to the injection of energy, momentum and mass into the ISM.  The origin of this energy is either massive stars and SNe (``stellar feedback") or supermassive black holes (``AGN feedback").  In terms of its impact on the ISM and any subsequent star formation activity, feedback can be either positive or negative.  Evidence for positive feedback includes for example the fact that in the Solar neighbourhood, almost all of the star formation happens on the surface of an outwardly-expanding bubble, triggered by a star formation event and ensuing SNe that took place near the bubble's centre approximately 14 million years ago \citep{zucker22}.  Conversely, negative feedback refers to the suppression of star formation activity through either the removal or heating of ISM gas.  For example, we infer that AGN feedback in early-type galaxies is crucial as a source of negative feedback. By heating the HIM, this feedback prevents the formation of cooling flows, which would otherwise be expected from theory, since the cooling timescale is shorter than the Hubble time.

\subsection{Connection to outflows}
\label{sec:SFconnection}

When feedback is powerful enough, it can not only impact the ISM in the immediate vicinity of the energy source, but also drive bulk motions of ISM gas; we describe ISM gas escaping its galaxy as an outflow. Outflows are the third important ingredient of the baryon cycle within the equilibrium model of galaxy evolution, in addition to inflows and star formation.  While stellar feedback operates in galaxies of all masses, it has a particularly large impact on the ISM of low-mass galaxies, since in these systems the typical velocity of SNe-driven winds is similar to the escape velocity of the system.  In massive galaxies, the dominant source of feedback that can power outflows is AGN.  Numerical simulations of galaxy formation implement stellar and AGN feedback in different ways, but regardless of these details, the consensus is that feedback and outflows are essential to produce galaxy populations that resemble those observed in the Universe today (in particular regarding the abundance of massive star-forming galaxies). 

The strength of the outflows and their potential for affecting the mass growth of galaxies through future star formation is usually parametrised by the mass-loading factor, the ratio of the mass outflow rate and the star formation rate ($\eta \equiv \dot{M}_{\rm out}/{\rm SFR}$).   If $\eta \gg 1$, then the galaxy loses ISM material to outflows at a significantly faster rate than it can covert it into stars, therefore leading to the suppression of star formation activity and in many cases, quenching.  Numerical simulations typically require outflows with significant mass-loading factors to achieve realistic outcomes.  While such high values of $\eta$ can be observed in highly star-forming or AGN-powered galaxies, it remains to be seen if such efficient outflows can be produced in more typical star-forming galaxies; this remains an active area of study. 

Outflows have been observed in multiple phases, from X-ray detected hot gas to cold molecular gas. For very actively star-forming galaxies, it is observed that the vast majority of the outflowing gas mass is molecular, with a contribution of atomic gas at the $\sim 10\%$ level, while ionised gas in the outflow is negligible in comparison \citep{fluetsch21}.  Once again, in the future it will be crucial to test wether the multi-phase nature of the outflowing material changes significantly depending on the energy source or the properties of the host galaxy. Other important open questions relate to the mechanisms through which energy is transferred from their source to the ISM gas, and the connection between the multiple phases of the outflowing material. The observation of dust within outflows could for example shed some light on the importance of radiation pressure on dust grains as compared e.g. to adiabatic expansion of gas powered by SNe.

\section{Conclusions}

The ISM plays a crucial role in driving galaxy evolution through the process of star formation. Observations of the ISM allow us to learn not only about star formation, but also about the other elements of the baryon cycle, namely the accretion of gas into the ISM, and the return of material to the external environment through outflows. Areas of particular activity in the coming years are likely to be the study of ISM chemistry and star formation in the earliest galaxies, driven by the ability of JWST to obtain detailed spectra of the very first galaxies. These observations are currently proving challenging to explain with current models, and therefore driving a burst of theoretical activity \citep[e.g.][]{dekel23}.  In the lower redshift Universe, a significant challenge still in front of us is to expand further our understanding of the ISM and star formation in low mass dwarf galaxies. This is a challenging topic, given the difficulty in probing their cold ISM through molecular emission lines or dust continuum, but an essential one as we work towards the ambitious objective of a complete understanding of the baryon cycle, in galaxies of all masses and across cosmic time. The ISM will remain one of the richest sources of information at our disposal in this endeavour.

\begin{ack}[Acknowledgments]

 Thank you to Michael Romano, Judith Sanders, Manasvee Saraf, Federico Speranza, and Niankun Yu for careful proof-reading of an advanced draft of this article, resulting in improved clarity of the text.  

\end{ack}

%\seealso{article title article title}

\bibliographystyle{Harvard}
%\bibliography{/Users/admin/Work/LaTeX/refs_amelie}

\begin{thebibliography*}{31}
\providecommand{\bibtype}[1]{}
\providecommand{\natexlab}[1]{#1}
{\catcode`\|=0\catcode`\#=12\catcode`\@=11\catcode`\\=12
|immediate|write|@auxout{\expandafter\ifx\csname
  natexlab\endcsname\relax\gdef\natexlab#1{#1}\fi}}
\renewcommand{\url}[1]{{\tt #1}}
\providecommand{\urlprefix}{URL }
\expandafter\ifx\csname urlstyle\endcsname\relax
  \providecommand{\doi}[1]{doi:\discretionary{}{}{}#1}\else
  \providecommand{\doi}{doi:\discretionary{}{}{}\begingroup
  \urlstyle{rm}\Url}\fi
\providecommand{\bibinfo}[2]{#2}
\providecommand{\eprint}[2][]{\url{#2}}

\bibtype{Article}%
\bibitem[{Agertz} et al.(2009)]{agertz09}
\bibinfo{author}{{Agertz} O}, \bibinfo{author}{{Teyssier} R} and
  \bibinfo{author}{{Moore} B} (\bibinfo{year}{2009}), \bibinfo{month}{Jul.}
\bibinfo{title}{{Disc formation and the origin of clumpy galaxies at high
  redshift}}.
\bibinfo{journal}{{\em \mnras}} \bibinfo{volume}{397} (\bibinfo{number}{1}):
  \bibinfo{pages}{L64--L68}.
  \bibinfo{doi}{\doi{10.1111/j.1745-3933.2009.00685.x}}.
\eprint{0901.2536}.

\bibtype{Article}%
\bibitem[{Arriagada-Neira} et al.(2024)]{arriagadaneira24}
\bibinfo{author}{{Arriagada-Neira} S}, \bibinfo{author}{{Herrera-Camus} R},
  \bibinfo{author}{{Villanueva} V}, \bibinfo{author}{{F{\"o}rster Schreiber}
  NM}, \bibinfo{author}{{Lee} M}, \bibinfo{author}{{Bolatto} A},
  \bibinfo{author}{{Chen} J}, \bibinfo{author}{{Genzel} R},
  \bibinfo{author}{{Liu} D}, \bibinfo{author}{{Renzini} A},
  \bibinfo{author}{{Tacconi} LJ}, \bibinfo{author}{{Tozzi} G} and
  \bibinfo{author}{{{\"U}bler} H} (\bibinfo{year}{2024}), \bibinfo{month}{Oct.}
\bibinfo{title}{{Deep kiloparsec view of the molecular gas in a massive
  star-forming galaxy at cosmic noon}}.
\bibinfo{journal}{{\em arXiv e-prints}} ,
  \bibinfo{eid}{arXiv:2410.14781}\bibinfo{doi}{\doi{10.48550/arXiv.2410.14781}}.
\eprint{2410.14781}.

\bibtype{Article}%
\bibitem[{Barnes} et al.(2001)]{HIPASS}
\bibinfo{author}{{Barnes} DG}, \bibinfo{author}{{Staveley-Smith} L},
  \bibinfo{author}{{de Blok} WJG}, \bibinfo{author}{{Oosterloo} T},
  \bibinfo{author}{{Stewart} IM}, \bibinfo{author}{{Wright} AE},
  \bibinfo{author}{{Banks} GD}, \bibinfo{author}{{Bhathal} R},
  \bibinfo{author}{{Boyce} PJ}, \bibinfo{author}{{Calabretta} MR},
  \bibinfo{author}{{Disney} MJ}, \bibinfo{author}{{Drinkwater} MJ},
  \bibinfo{author}{{Ekers} RD}, \bibinfo{author}{{Freeman} KC},
  \bibinfo{author}{{Gibson} BK}, \bibinfo{author}{{Green} AJ},
  \bibinfo{author}{{Haynes} RF}, \bibinfo{author}{{te Lintel Hekkert} P},
  \bibinfo{author}{{Henning} PA}, \bibinfo{author}{{Jerjen} H},
  \bibinfo{author}{{Juraszek} S}, \bibinfo{author}{{Kesteven} MJ},
  \bibinfo{author}{{Kilborn} VA}, \bibinfo{author}{{Knezek} PM},
  \bibinfo{author}{{Koribalski} B}, \bibinfo{author}{{Kraan-Korteweg} RC},
  \bibinfo{author}{{Malin} DF}, \bibinfo{author}{{Marquarding} M},
  \bibinfo{author}{{Minchin} RF}, \bibinfo{author}{{Mould} JR},
  \bibinfo{author}{{Price} RM}, \bibinfo{author}{{Putman} ME},
  \bibinfo{author}{{Ryder} SD}, \bibinfo{author}{{Sadler} EM},
  \bibinfo{author}{{Schr{\"o}der} A}, \bibinfo{author}{{Stootman} F},
  \bibinfo{author}{{Webster} RL}, \bibinfo{author}{{Wilson} WE} and
  \bibinfo{author}{{Ye} T} (\bibinfo{year}{2001}), \bibinfo{month}{Apr.}
\bibinfo{title}{{The Hi Parkes All Sky Survey: southern observations,
  calibration and robust imaging}}.
\bibinfo{journal}{{\em \mnras}} \bibinfo{volume}{322}:
  \bibinfo{pages}{486--498}.
  \bibinfo{doi}{\doi{10.1046/j.1365-8711.2001.04102.x}}.

\bibtype{Article}%
\bibitem[{Borthakur} et al.(2024)]{borthakur24}
\bibinfo{author}{{Borthakur} S}, \bibinfo{author}{{Padave} M},
  \bibinfo{author}{{Heckman} T}, \bibinfo{author}{{Gim} HB},
  \bibinfo{author}{{Olvera} AJ}, \bibinfo{author}{{Koplitz} B},
  \bibinfo{author}{{Momjian} E}, \bibinfo{author}{{Jansen} RA},
  \bibinfo{author}{{Thilker} D}, \bibinfo{author}{{Kauffman} G},
  \bibinfo{author}{{Fox} AJ}, \bibinfo{author}{{Tumlinson} J},
  \bibinfo{author}{{Kennicutt} RC}, \bibinfo{author}{{Nelson} D},
  \bibinfo{author}{{Monckiewicz} J} and  \bibinfo{author}{{Naab} T}
  (\bibinfo{year}{2024}), \bibinfo{month}{Sep.}
\bibinfo{title}{{DIISC Survey: Deciphering the Interplay Between the
  Interstellar Medium, Stars, and the Circumgalactic Medium Survey}}.
\bibinfo{journal}{{\em arXiv e-prints}} ,
  \bibinfo{eid}{arXiv:2409.12554}\bibinfo{doi}{\doi{10.48550/arXiv.2409.12554}}.
\eprint{2409.12554}.

\bibtype{Article}%
\bibitem[{Chevance} et al.(2020)]{chevance20review}
\bibinfo{author}{{Chevance} M}, \bibinfo{author}{{Kruijssen} JMD},
  \bibinfo{author}{{Vazquez-Semadeni} E}, \bibinfo{author}{{Nakamura} F},
  \bibinfo{author}{{Klessen} R}, \bibinfo{author}{{Ballesteros-Paredes} J},
  \bibinfo{author}{{Inutsuka} Si}, \bibinfo{author}{{Adamo} A} and
  \bibinfo{author}{{Hennebelle} P} (\bibinfo{year}{2020}),
  \bibinfo{month}{Apr.}
\bibinfo{title}{{The Molecular Cloud Lifecycle}}.
\bibinfo{journal}{{\em \ssr}} \bibinfo{volume}{216} (\bibinfo{number}{4}),
  \bibinfo{eid}{50}. \bibinfo{doi}{\doi{10.1007/s11214-020-00674-x}}.
\eprint{2004.06113}.

\bibtype{Article}%
\bibitem[{Cortese} et al.(2021)]{cortese21}
\bibinfo{author}{{Cortese} L}, \bibinfo{author}{{Catinella} B} and
  \bibinfo{author}{{Smith} R} (\bibinfo{year}{2021}), \bibinfo{month}{Aug.}
\bibinfo{title}{{The Dawes Review 9: The role of cold gas stripping on the star
  formation quenching of satellite galaxies}}.
\bibinfo{journal}{{\em \pasa}} \bibinfo{volume}{38}, \bibinfo{eid}{e035}.
  \bibinfo{doi}{\doi{10.1017/pasa.2021.18}}.
\eprint{2104.02193}.

\bibtype{Article}%
\bibitem[{Davis} et al.(2022)]{davis22}
\bibinfo{author}{{Davis} TA}, \bibinfo{author}{{Gensior} J},
  \bibinfo{author}{{Bureau} M}, \bibinfo{author}{{Cappellari} M},
  \bibinfo{author}{{Choi} W}, \bibinfo{author}{{Elford} JS},
  \bibinfo{author}{{Kruijssen} JMD}, \bibinfo{author}{{Lelli} F},
  \bibinfo{author}{{Liang} FH}, \bibinfo{author}{{Liu} L},
  \bibinfo{author}{{Ruffa} I}, \bibinfo{author}{{Saito} T},
  \bibinfo{author}{{Sarzi} M}, \bibinfo{author}{{Schruba} A} and
  \bibinfo{author}{{Williams} TG} (\bibinfo{year}{2022}), \bibinfo{month}{May}.
\bibinfo{title}{{WISDOM Project - X. The morphology of the molecular ISM in
  galaxy centres and its dependence on galaxy structure}}.
\bibinfo{journal}{{\em \mnras}} \bibinfo{volume}{512} (\bibinfo{number}{1}):
  \bibinfo{pages}{1522--1540}. \bibinfo{doi}{\doi{10.1093/mnras/stac600}}.
\eprint{2203.01358}.

\bibtype{Article}%
\bibitem[{Dekel} et al.(2023)]{dekel23}
\bibinfo{author}{{Dekel} A}, \bibinfo{author}{{Sarkar} KC},
  \bibinfo{author}{{Birnboim} Y}, \bibinfo{author}{{Mandelker} N} and
  \bibinfo{author}{{Li} Z} (\bibinfo{year}{2023}), \bibinfo{month}{Aug.}
\bibinfo{title}{{Efficient formation of massive galaxies at cosmic dawn by
  feedback-free starbursts}}.
\bibinfo{journal}{{\em \mnras}} \bibinfo{volume}{523} (\bibinfo{number}{3}):
  \bibinfo{pages}{3201--3218}. \bibinfo{doi}{\doi{10.1093/mnras/stad1557}}.
\eprint{2303.04827}.

\bibtype{Article}%
\bibitem[{Feldmann} et al.(2023)]{feldmann23}
\bibinfo{author}{{Feldmann} R}, \bibinfo{author}{{Quataert} E},
  \bibinfo{author}{{Faucher-Gigu{\`e}re} CA}, \bibinfo{author}{{Hopkins} PF},
  \bibinfo{author}{{{\c{C}}atmabacak} O}, \bibinfo{author}{{Kere{\v{s}}} D},
  \bibinfo{author}{{Bassini} L}, \bibinfo{author}{{Bernardini} M},
  \bibinfo{author}{{Bullock} JS}, \bibinfo{author}{{Cenci} E},
  \bibinfo{author}{{Gensior} J}, \bibinfo{author}{{Liang} L},
  \bibinfo{author}{{Moreno} J} and  \bibinfo{author}{{Wetzel} A}
  (\bibinfo{year}{2023}), \bibinfo{month}{Jul.}
\bibinfo{title}{{FIREbox: simulating galaxies at high dynamic range in a
  cosmological volume}}.
\bibinfo{journal}{{\em \mnras}} \bibinfo{volume}{522} (\bibinfo{number}{3}):
  \bibinfo{pages}{3831--3860}. \bibinfo{doi}{\doi{10.1093/mnras/stad1205}}.
\eprint{2205.15325}.

\bibtype{Article}%
\bibitem[{Fluetsch} et al.(2021)]{fluetsch21}
\bibinfo{author}{{Fluetsch} A}, \bibinfo{author}{{Maiolino} R},
  \bibinfo{author}{{Carniani} S}, \bibinfo{author}{{Arribas} S},
  \bibinfo{author}{{Belfiore} F}, \bibinfo{author}{{Bellocchi} E},
  \bibinfo{author}{{Cazzoli} S}, \bibinfo{author}{{Cicone} C},
  \bibinfo{author}{{Cresci} G}, \bibinfo{author}{{Fabian} AC},
  \bibinfo{author}{{Gallagher} R}, \bibinfo{author}{{Ishibashi} W},
  \bibinfo{author}{{Mannucci} F}, \bibinfo{author}{{Marconi} A},
  \bibinfo{author}{{Perna} M}, \bibinfo{author}{{Sturm} E} and
  \bibinfo{author}{{Venturi} G} (\bibinfo{year}{2021}), \bibinfo{month}{Aug.}
\bibinfo{title}{{Properties of the multiphase outflows in local (ultra)luminous
  infrared galaxies}}.
\bibinfo{journal}{{\em \mnras}} \bibinfo{volume}{505} (\bibinfo{number}{4}):
  \bibinfo{pages}{5753--5783}. \bibinfo{doi}{\doi{10.1093/mnras/stab1666}}.
\eprint{2006.13232}.

\bibtype{Article}%
\bibitem[{F{\"o}rster Schreiber} and {Wuyts}(2020)]{forster20}
\bibinfo{author}{{F{\"o}rster Schreiber} NM} and  \bibinfo{author}{{Wuyts} S}
  (\bibinfo{year}{2020}), \bibinfo{month}{Aug.}
\bibinfo{title}{{Star-Forming Galaxies at Cosmic Noon}}.
\bibinfo{journal}{{\em \araa}} \bibinfo{volume}{58}: \bibinfo{pages}{661--725}.
  \bibinfo{doi}{\doi{10.1146/annurev-astro-032620-021910}}.
\eprint{2010.10171}.

\bibtype{Inproceedings}%
\bibitem[{Fraternali}(2017)]{fraternali17}
\bibinfo{author}{{Fraternali} F} (\bibinfo{year}{2017}), \bibinfo{month}{Jan.},
  \bibinfo{title}{{Gas Accretion via Condensation and Fountains}},
  \bibinfo{editor}{{Fox} A} and  \bibinfo{editor}{{Dav{\'e}} R}, (Eds.),
  \bibinfo{booktitle}{Gas Accretion onto Galaxies},
  \bibinfo{series}{Astrophysics and Space Science Library},
  \bibinfo{volume}{430}, pp. \bibinfo{pages}{323}, \eprint{1612.00477}.

\bibtype{Article}%
\bibitem[{Giovanelli} et al.(2005)]{alfalfa1}
\bibinfo{author}{{Giovanelli} R}, \bibinfo{author}{{Haynes} MP},
  \bibinfo{author}{{Kent} BR}, \bibinfo{author}{{Perillat} P},
  \bibinfo{author}{{Saintonge} A}, \bibinfo{author}{{Brosch} N},
  \bibinfo{author}{{Catinella} B} and  \bibinfo{author}{et~al.}
  (\bibinfo{year}{2005}), \bibinfo{month}{Dec.}
\bibinfo{title}{{The Arecibo Legacy Fast ALFA Survey. I. Science Goals, Survey
  Design, and Strategy}}.
\bibinfo{journal}{{\em \aj}} \bibinfo{volume}{130}:
  \bibinfo{pages}{2598--2612}. \bibinfo{doi}{\doi{10.1086/497431}}.
\eprint{arXiv:astro-ph/0508301}.

\bibtype{Article}%
\bibitem[{Haffner} et al.(2009)]{haffner09}
\bibinfo{author}{{Haffner} LM}, \bibinfo{author}{{Dettmar} RJ},
  \bibinfo{author}{{Beckman} JE}, \bibinfo{author}{{Wood} K},
  \bibinfo{author}{{Slavin} JD}, \bibinfo{author}{{Giammanco} C},
  \bibinfo{author}{{Madsen} GJ}, \bibinfo{author}{{Zurita} A} and
  \bibinfo{author}{{Reynolds} RJ} (\bibinfo{year}{2009}), \bibinfo{month}{Jul.}
\bibinfo{title}{{The warm ionized medium in spiral galaxies}}.
\bibinfo{journal}{{\em Reviews of Modern Physics}} \bibinfo{volume}{81}
  (\bibinfo{number}{3}): \bibinfo{pages}{969--997}.
  \bibinfo{doi}{\doi{10.1103/RevModPhys.81.969}}.
\eprint{0901.0941}.

\bibtype{Article}%
\bibitem[{Hopkins} et al.(2018)]{hopkins18}
\bibinfo{author}{{Hopkins} PF}, \bibinfo{author}{{Wetzel} A},
  \bibinfo{author}{{Kere{\v{s}}} D}, \bibinfo{author}{{Faucher-Gigu{\`e}re}
  CA}, \bibinfo{author}{{Quataert} E}, \bibinfo{author}{{Boylan-Kolchin} M},
  \bibinfo{author}{{Murray} N}, \bibinfo{author}{{Hayward} CC},
  \bibinfo{author}{{Garrison-Kimmel} S}, \bibinfo{author}{{Hummels} C},
  \bibinfo{author}{{Feldmann} R}, \bibinfo{author}{{Torrey} P},
  \bibinfo{author}{{Ma} X}, \bibinfo{author}{{Angl{\'e}s-Alc{\'a}zar} D},
  \bibinfo{author}{{Su} KY}, \bibinfo{author}{{Orr} M},
  \bibinfo{author}{{Schmitz} D}, \bibinfo{author}{{Escala} I},
  \bibinfo{author}{{Sanderson} R}, \bibinfo{author}{{Grudi{\'c}} MY},
  \bibinfo{author}{{Hafen} Z}, \bibinfo{author}{{Kim} JH},
  \bibinfo{author}{{Fitts} A}, \bibinfo{author}{{Bullock} JS},
  \bibinfo{author}{{Wheeler} C}, \bibinfo{author}{{Chan} TK},
  \bibinfo{author}{{Elbert} OD} and  \bibinfo{author}{{Narayanan} D}
  (\bibinfo{year}{2018}), \bibinfo{month}{Oct.}
\bibinfo{title}{{FIRE-2 simulations: physics versus numerics in galaxy
  formation}}.
\bibinfo{journal}{{\em \mnras}} \bibinfo{volume}{480} (\bibinfo{number}{1}):
  \bibinfo{pages}{800--863}. \bibinfo{doi}{\doi{10.1093/mnras/sty1690}}.
\eprint{1702.06148}.

\bibtype{Article}%
\bibitem[{Hunter} et al.(2024)]{hunter24}
\bibinfo{author}{{Hunter} DA}, \bibinfo{author}{{Elmegreen} BG} and
  \bibinfo{author}{{Madden} SC} (\bibinfo{year}{2024}), \bibinfo{month}{Sep.}
\bibinfo{title}{{The Interstellar Medium in Dwarf Irregular Galaxies}}.
\bibinfo{journal}{{\em \araa}} \bibinfo{volume}{62} (\bibinfo{number}{1}):
  \bibinfo{pages}{113--155}.
  \bibinfo{doi}{\doi{10.1146/annurev-astro-052722-104109}}.
\eprint{2402.17004}.

\bibtype{Article}%
\bibitem[{Kere{\v s}} et al.(2005)]{keres05}
\bibinfo{author}{{Kere{\v s}} D}, \bibinfo{author}{{Katz} N},
  \bibinfo{author}{{Weinberg} DH} and  \bibinfo{author}{{Dav{\'e}} R}
  (\bibinfo{year}{2005}), \bibinfo{month}{Oct.}
\bibinfo{title}{{How do galaxies get their gas?}}
\bibinfo{journal}{{\em \mnras}} \bibinfo{volume}{363}: \bibinfo{pages}{2--28}.
  \bibinfo{doi}{\doi{10.1111/j.1365-2966.2005.09451.x}}.
\eprint{arXiv:astro-ph/0407095}.

\bibtype{Article}%
\bibitem[{Lilly} et al.(2013)]{lilly13}
\bibinfo{author}{{Lilly} SJ}, \bibinfo{author}{{Carollo} CM},
  \bibinfo{author}{{Pipino} A}, \bibinfo{author}{{Renzini} A} and
  \bibinfo{author}{{Peng} Y} (\bibinfo{year}{2013}), \bibinfo{month}{Aug.}
\bibinfo{title}{{Gas Regulation of Galaxies: The Evolution of the Cosmic
  Specific Star Formation Rate, the Metallicity-Mass-Star-formation Rate
  Relation, and the Stellar Content of Halos}}.
\bibinfo{journal}{{\em \apj}} \bibinfo{volume}{772}, \bibinfo{eid}{119}.
  \bibinfo{doi}{\doi{10.1088/0004-637X/772/2/119}}.
\eprint{1303.5059}.

\bibtype{Article}%
\bibitem[{Madau} and {Dickinson}(2014)]{madaudickinson14}
\bibinfo{author}{{Madau} P} and  \bibinfo{author}{{Dickinson} M}
  (\bibinfo{year}{2014}), \bibinfo{month}{Aug.}
\bibinfo{title}{{Cosmic Star-Formation History}}.
\bibinfo{journal}{{\em \araa}} \bibinfo{volume}{52}: \bibinfo{pages}{415--486}.
  \bibinfo{doi}{\doi{10.1146/annurev-astro-081811-125615}}.
\eprint{1403.0007}.

\bibtype{Article}%
\bibitem[{Martig} et al.(2009)]{martig09}
\bibinfo{author}{{Martig} M}, \bibinfo{author}{{Bournaud} F},
  \bibinfo{author}{{Teyssier} R} and  \bibinfo{author}{{Dekel} A}
  (\bibinfo{year}{2009}), \bibinfo{month}{Dec.}
\bibinfo{title}{{Morphological Quenching of Star Formation: Making Early-Type
  Galaxies Red}}.
\bibinfo{journal}{{\em \apj}} \bibinfo{volume}{707}: \bibinfo{pages}{250--267}.
  \bibinfo{doi}{\doi{10.1088/0004-637X/707/1/250}}.
\eprint{0905.4669}.

\bibtype{Article}%
\bibitem[{McKee} and {Ostriker}(1977)]{mckee77}
\bibinfo{author}{{McKee} CF} and  \bibinfo{author}{{Ostriker} JP}
  (\bibinfo{year}{1977}), \bibinfo{month}{Nov.}
\bibinfo{title}{{A theory of the interstellar medium: three components
  regulated by supernova explosions in an inhomogeneous substrate.}}
\bibinfo{journal}{{\em \apj}} \bibinfo{volume}{218}: \bibinfo{pages}{148--169}.
  \bibinfo{doi}{\doi{10.1086/155667}}.

\bibtype{Book}%
\bibitem[Ryden and Pogge(2021)]{ryden21}
\bibinfo{author}{Ryden B} and  \bibinfo{author}{Pogge RW}
  (\bibinfo{year}{2021}).
\bibinfo{title}{{Interstellar and Intergalactic Medium}},
  \bibinfo{publisher}{Cambridge University Press}.

\bibtype{Article}%
\bibitem[{Saintonge} and {Catinella}(2022)]{saintonge22}
\bibinfo{author}{{Saintonge} A} and  \bibinfo{author}{{Catinella} B}
  (\bibinfo{year}{2022}), \bibinfo{month}{Aug.}
\bibinfo{title}{{The Cold Interstellar Medium of Galaxies in the Local
  Universe}}.
\bibinfo{journal}{{\em \araa}} \bibinfo{volume}{60}: \bibinfo{pages}{319--361}.
  \bibinfo{doi}{\doi{10.1146/annurev-astro-021022-043545}}.
\eprint{2202.00690}.

\bibtype{Article}%
\bibitem[{Saintonge} et al.(2017)]{saintonge17}
\bibinfo{author}{{Saintonge} A}, \bibinfo{author}{{Catinella} B},
  \bibinfo{author}{{Tacconi} LJ}, \bibinfo{author}{{Kauffmann} G},
  \bibinfo{author}{{Genzel} R}, \bibinfo{author}{{Cortese} L},
  \bibinfo{author}{{Dav{\'e}} R}, \bibinfo{author}{{Fletcher} TJ} and
  \bibinfo{author}{et~al.} (\bibinfo{year}{2017}), \bibinfo{month}{Dec.}
\bibinfo{title}{{xCOLD GASS: The Complete IRAM 30 m Legacy Survey of Molecular
  Gas for Galaxy Evolution Studies}}.
\bibinfo{journal}{{\em \apjs}} \bibinfo{volume}{233}, \bibinfo{eid}{22}.
  \bibinfo{doi}{\doi{10.3847/1538-4365/aa97e0}}.
\eprint{1710.02157}.

\bibtype{Article}%
\bibitem[{Schaye} et al.(2015)]{schaye15}
\bibinfo{author}{{Schaye} J}, \bibinfo{author}{{Crain} RA},
  \bibinfo{author}{{Bower} RG}, \bibinfo{author}{{Furlong} M},
  \bibinfo{author}{{Schaller} M}, \bibinfo{author}{{Theuns} T},
  \bibinfo{author}{{Dalla Vecchia} C}, \bibinfo{author}{{Frenk} CS},
  \bibinfo{author}{{McCarthy} IG}, \bibinfo{author}{{Helly} JC},
  \bibinfo{author}{{Jenkins} A}, \bibinfo{author}{{Rosas-Guevara} YM},
  \bibinfo{author}{{White} SDM}, \bibinfo{author}{{Baes} M},
  \bibinfo{author}{{Booth} CM}, \bibinfo{author}{{Camps} P},
  \bibinfo{author}{{Navarro} JF}, \bibinfo{author}{{Qu} Y},
  \bibinfo{author}{{Rahmati} A}, \bibinfo{author}{{Sawala} T},
  \bibinfo{author}{{Thomas} PA} and  \bibinfo{author}{{Trayford} J}
  (\bibinfo{year}{2015}), \bibinfo{month}{Jan.}
\bibinfo{title}{{The EAGLE project: simulating the evolution and assembly of
  galaxies and their environments}}.
\bibinfo{journal}{{\em \mnras}} \bibinfo{volume}{446} (\bibinfo{number}{1}):
  \bibinfo{pages}{521--554}. \bibinfo{doi}{\doi{10.1093/mnras/stu2058}}.
\eprint{1407.7040}.

\bibtype{Article}%
\bibitem[{Spitzer}(1956)]{spitzer56}
\bibinfo{author}{{Spitzer} Lyman J} (\bibinfo{year}{1956}),
  \bibinfo{month}{Jul.}
\bibinfo{title}{{On a Possible Interstellar Galactic Corona.}}
\bibinfo{journal}{{\em \apj}} \bibinfo{volume}{124}: \bibinfo{pages}{20}.
  \bibinfo{doi}{\doi{10.1086/146200}}.

\bibtype{Article}%
\bibitem[{Tacconi} et al.(2020)]{tacconi20}
\bibinfo{author}{{Tacconi} LJ}, \bibinfo{author}{{Genzel} R} and
  \bibinfo{author}{{Sternberg} A} (\bibinfo{year}{2020}), \bibinfo{month}{Aug.}
\bibinfo{title}{{The Evolution of the Star-Forming Interstellar Medium Across
  Cosmic Time}}.
\bibinfo{journal}{{\em \araa}} \bibinfo{volume}{58}: \bibinfo{pages}{157--203}.
  \bibinfo{doi}{\doi{10.1146/annurev-astro-082812-141034}}.
\eprint{2003.06245}.

\bibtype{Book}%
\bibitem[{Tielens}(2005)]{tielens05}
\bibinfo{author}{{Tielens} AGGM} (\bibinfo{year}{2005}).
\bibinfo{title}{{The Physics and Chemistry of the Interstellar Medium}}.

\bibtype{Article}%
\bibitem[{Tumlinson} et al.(2017)]{tumlinson17}
\bibinfo{author}{{Tumlinson} J}, \bibinfo{author}{{Peeples} MS} and
  \bibinfo{author}{{Werk} JK} (\bibinfo{year}{2017}), \bibinfo{month}{Aug.}
\bibinfo{title}{{The Circumgalactic Medium}}.
\bibinfo{journal}{{\em \araa}} \bibinfo{volume}{55} (\bibinfo{number}{1}):
  \bibinfo{pages}{389--432}.
  \bibinfo{doi}{\doi{10.1146/annurev-astro-091916-055240}}.
\eprint{1709.09180}.

\bibtype{Article}%
\bibitem[{Wolfire} et al.(2003)]{wolfire03}
\bibinfo{author}{{Wolfire} MG}, \bibinfo{author}{{McKee} CF},
  \bibinfo{author}{{Hollenbach} D} and  \bibinfo{author}{{Tielens} AGGM}
  (\bibinfo{year}{2003}), \bibinfo{month}{Apr.}
\bibinfo{title}{{Neutral Atomic Phases of the Interstellar Medium in the
  Galaxy}}.
\bibinfo{journal}{{\em \apj}} \bibinfo{volume}{587} (\bibinfo{number}{1}):
  \bibinfo{pages}{278--311}. \bibinfo{doi}{\doi{10.1086/368016}}.
\eprint{astro-ph/0207098}.

\bibtype{Article}%
\bibitem[{Zucker} et al.(2022)]{zucker22}
\bibinfo{author}{{Zucker} C}, \bibinfo{author}{{Goodman} AA},
  \bibinfo{author}{{Alves} J}, \bibinfo{author}{{Bialy} S},
  \bibinfo{author}{{Foley} M}, \bibinfo{author}{{Speagle} JS},
  \bibinfo{author}{{Gro{\^I}{\texttwosuperior}schedl} J},
  \bibinfo{author}{{Finkbeiner} DP}, \bibinfo{author}{{Burkert} A},
  \bibinfo{author}{{Khimey} D} and  \bibinfo{author}{{Swiggum} C}
  (\bibinfo{year}{2022}), \bibinfo{month}{Jan.}
\bibinfo{title}{{Star formation near the Sun is driven by expansion of the
  Local Bubble}}.
\bibinfo{journal}{{\em \nat}} \bibinfo{volume}{601} (\bibinfo{number}{7893}):
  \bibinfo{pages}{334--337}. \bibinfo{doi}{\doi{10.1038/s41586-021-04286-5}}.
\eprint{2201.05124}.

\end{thebibliography*}

\end{document}